\documentclass[twocolumn]{bmcart}
\usepackage{float}
\usepackage{amsthm,amsmath}
\usepackage{hyperref}
\hypersetup{colorlinks,allcolors=black}
\usepackage[utf8]{inputenc} 
\usepackage{graphicx}

\usepackage{subcaption}

\usepackage{graphicx}
\usepackage{lscape}
\usepackage{rotating}
\usepackage[toc,page]{appendix}

\startlocaldefs
\endlocaldefs

\begin{document}

\begin{frontmatter}

\begin{fmbox}
\dochead{Research}

\title{Where are the vulnerable children? Identification and comparison of clusters of young children with health and developmental vulnerabilities across Queensland }

 \author[
  addressref={aff1},                   
 corref={aff1},                       
   email={w.areed@qut.edu.au}   
 ]
 {\inits{}\fnm{Wala} \snm{Draidi Areed}}
 \author[
   addressref={aff1},
   email={a11.price@qut.edu.au}
 ]{\inits{}\fnm{Aiden} \snm{Price}}
 \author[
   addressref={aff2},
   email={kathryn.arnett@health.qld.gov.au}
 ]{\inits{}\fnm{Kathryn} \snm{Arnett}}
 \author[
  addressref={aff1},
   email={k.mengersen@qut.edu.au}
 ]{\inits{}\fnm{Kerrie} \snm{Mengersen}}
 \author[
   addressref={aff1},
   email={helen.thompson@qut.edu.au}
 ]{\inits{}\fnm{Helen} \snm{Thompson}}


\address[id=aff1]{
   \orgdiv{School of Mathematical Science, Center for Data Science},             
   \orgname{Queensland University of Technology},          
   \city{Queensland},                              
   \cny{Australia}                                    
 }
 \address[id=aff2]{%
   \orgdiv{Children's Health Queensland },
   \street{},
   \postcode{}
   \city{Queensland},
   \cny{Australia}
 }





\begin{abstractbox}

\begin{abstract} 
\parttitle{Background} 
Developmental vulnerabilities within children in Queensland have a variety of domains; these domains measure the development of children in their first five years. It is crucial to understand how these domains are grouped, or clustered, with respect to population risk factor profiles. These groups inform policy implementation, which can help to provide assistance to the most vulnerable children across Queensland. 

\parttitle{Methods} 
$K$-means analysis was conducted on data from the Australian Early Development Census and the Australian Bureau of Statistics. The clusters were then compared with respect to their geographic locations and risk factor profiles. The results are presented in this paper and are publicly available via an interactive dashboard application in R Shiny.
\parttitle{Results} 
This study presents a comprehensive clustering analysis for child development vulnerability domains in Queensland. In addition, all of the clustering analyses reveal a strong relationship between developmentally vulnerable and socio-economic and remoteness factors. In addition, we found that children who attend preschool and whose primary language is English are, in most cases, in the lowest developmentally vulnerable cluster. 
\parttitle{Conclusion}
In this study, the performance of the K-means clustering algorithm has been developed to study the clusters inside child development vulnerabilities when analysing the data at the small area level. Further, R shiny application was created, and the feature of the risk factors in each region was studied. 
\end{abstract}




\end{abstractbox}
\end{fmbox} 

\end{frontmatter}



\section*{Background}
Internationally, there is an increasing focus on population health among integrated care organisations and health systems  \cite{noble2013can,sox2013resolving}. The goal of population health methods is to enhance the overall health of a group of people.
In order to do this, it is critical to recognise the requirements of various groups within the population \cite{alderwick2015population,porter2013redesigning,vuik2016quantitative}. An important group in the population is children. \\
Healthy child development improves human capabilities by allowing children to mature and participate in economic, social, and civic life \cite{zubrick2005resources}.  Child development includes the  biological, psychological, and emotional changes that occur between birth and maturity \cite{unicef1993facts}. Physical, social, emotional, speech and language, and communication skills are the five critical domains of growth \cite{irwin2007early}.
Children's development in the early years from birth to five years of age is crucial since it is at this time that the foundations for health development, emotional well-being, and life success are built \cite{hertzman1996child}.\\
Increasingly many countries, including Australia, are using national progress indicators of early childhood development to track critical developmental domains in the early years \cite{brinkman2014data,goldfeld2009process}. The Australian Early Development Census (AEDC) provides a nationwide snapshot of children's development at the time children commence their first year of full-time school, and reports scores across the five domains of growth. Each child is given a score between zero and ten for each of the AEDC domains and, using the cut-offs established as a baseline in 2009, children falling below the 10th percentile in a domain, taking into account the age differences, are categorised as `developmentally vulnerable'. The AEDC reveals that the proportion of children who are developmentally vulnerable, within each developmental domain, varies considerably between geographical regions across Australia. This variation exists across the smallest geographic areas defined by the AEDC, which are referred to as local communities and are often equivalent to suburbs \cite{snapshot}. To address inequalities in developmental vulnerabilities, further insight is needed into the factors that contribute to such variation \cite{lynch2010inequalities}. One method to understand the variation is cluster analysis \cite{peeters1989hierarchical}.\\
Cluster analysis is a mechanism for grouping (clustering) a set of objects (e.g., local communities) in such a way that objects within a group (cluster) are more similar (e.g., in terms of developmentally vulnerable) to one another than to those in other groups (clusters) \cite{zhai2014k}.
There are many clustering methods: model-based versus fully empirical, parametric versus non-parametric, probabilistic versus non-probabilistic, hierarchical versus partition-based, and supervised versus unsupervised \cite{sisodia2012clustering}. There are also many computational methods for  clustering, including the Expectation-Maximisation (EM) algorithm \cite{dempster1977maximum} and a variety of simulation-based algorithms, such as Markov chain Monte Carlo (MCMC) \cite{fruhwirth2019handbook}. A well-established simple non-probabilistic unsupervised partitioning method, which is employed in this study, is $K$-means clustering, where $K$ denotes the number of clusters (Section \ref{kmean}) \cite{venkatesan2015performance,yadav2013review}. Common strategies for choosing the value of $K$ include the elbow method \cite{bholowalia2014ebk}, gap statistic \cite{tibshirani2001estimating}, silhouette coefficient \cite{kaufman2009finding}, and canopy method \cite{yu2014research}.
In this study, the silhouette coefficient (Section \ref{silh}) is used to determine the number of clusters. While the elbow method is easy to implement and the calculations required are simple, the silhouette coefficient allows evaluations of clusters on multiple criteria, and hence it is more likely that the optimal number of clusters can be determined  \cite{ogbuabor2018clustering}.\\
 Publicly accessible data in the population AEDC domain are frequently aggregated within geographical areas \cite{buchin2008clusters}. In Australia, these geographical areas are typically the statistical areas defined in the Australian Statistical Geography Standard (ASGS). In the ASGS, Statistical Areas Level 1 (SA1) are the smallest defined geographical areas and aggregate to form Statistical Areas Level 2 (SA2). There are four levels of aggregation of statistical areas, SA1 through SA4. Where personal-level information is available, it is not uncommon for data on the exact location of individuals to be missing. Even if exact location data are available, privacy and confidentiality concerns prevent publication of person-level information. Hence this study uses data aggregated at SA2 level \cite{buchin2008clusters}.\\
 In this study, we aimed to identify and characterise regions in Queensland in terms of high and low vulnerability across five domains of health and  development, for children in their first year of full time school. We used $K$-means clustering to identify regions, such that within a cluster of SA2s making up a region, children have similar vulnerabilities for a given AEDC domain. In characterising the regions (clusters of SA2s), we consider the factors from AEDC  which are publicly available in SA2 level: attendance at preschool, Indigenous status, mother's language, country of birth, socioeconomic status, and remoteness status.
 In addition to the abridged results presented in this paper, we developed a web application to make the complete set of results accessible and more easily digestible. The web application has an intuitive interface that allows users to interactively explore child development vulnerability across the AEDC domains and across the SA2 areas of Queensland. We used the Shiny package for R \cite{chang2015package} to develop the web application, since the data were analysed also using R statistical software \cite{r}.
 The results of this research will support targeted early intervention strategies which can allow children to reach their maximum developmental potential.

\section*{Methods}

\subsection*{Case study and sources of data} \label{mat}
Child development vulnerability data were obtained from the 2018 Australian Early Development Census (AEDC). The AEDC is conducted every three years and collects data on children in their first year of full time school. The AEDC recently took place in 2021, but the most recent data available is for the 2018 census. The census is performed by classroom teachers in the child’s first year of full-time schooling across Australian Government and non-Government schools, and data are collected with the agreement of parents \cite{snapshot}. The data provided on a child by their teacher, based on the teacher's knowledge and observations of the child, is used to assign the child a score (0 to 10) for each AEDC developmental domain . For each domain, the child is then classified as \emph{vulnerable} if their score is in the lowest 10\% of scores for that domain using the cut-offs established as a baseline in 2009. Approximately 65,000 children (98.1\% of eligible children) across 1,414 Queensland Government, Catholic and Independent schools participated in the 2018 AEDC collection. The data were available as aggregated counts at the SA2 level. Among the 528 SA2s that make up Queensland, there was an average of 123 children per SA2, with a standard deviation of 100 \cite{snapshot}.\\
All five domains of health and developmentally vulnerable from the AEDC were considered in this study: physical health and well-being (Physical), social competence (Social), emotional maturity (Emotional), language and cognitive skills-school based (Language), and communication skills and general knowledge (Communication).
We also considered two additional AEDC indicators of vulnerability: vulnerable in one or more domain (Vuln 1), and vulnerable in two or more domains (Vuln 2). 
Due to the aggregated nature of the available data, we focused on the proportion of vulnerable children within each SA2. The following data were also obtained from the 2018 AEDC for each SA2: proportion of children who attended pre-school (Preschool), proportion who identified as Indigenous (Indigenous), proportion with English as the mother's language (English), proportion with Australia as country of birth (Australia). Further data for 2018 were obtained from the Australian Bureau of Statistics (ABS) for each SA2 including: Index of Relative Socio-economic Disadvantage (IRSD) for the SA2 (1 to 10), and Remoteness (Major City, Inner Regional, Outer Regional, Remote, Very Remote). The IRSD is coded from 1 (lowest) to 10 (highest) \cite{walker2005index}; a low score suggests that the area, in general, is at a disadvantage, e.g., many low-income households, many people without qualifications or with low-skill occupations. In 2018, there were 294 major city, 113 inner regional, 96 outer regional, 11 remote and 14 very remote SA2s in Queensland. \\
Between 3\% and 6\% of the data were missing variables in the dataset. Proportions (e.g., Preschool, Indigenous) that were missing for an SA2 were imputed using the average of the proportions from the neighbouring SA2s. For categorical data, i.e; IRSD and Remoteness,  the missing value was imputed using the highest frequency category of the neighbourhood SA2s. Missing values for two islands could not be imputed, as the regions have no contiguous neighbours. As a result, the analysis carried out in this study was reduced to the remaining 526 SA2 areas.
\subsection*{Clustering method} \label{method}
This section details the clustering method used to investigate the data clusters. All statistical analyses were conducted using R statistical software version R-4.1.3 \cite{r}. The analyses for the $K$-means algorithm were carried out using  mclust \cite{fraley2012package}, and factoextra \cite{kassambara2017package} packages, and the shiny package in R was used to develop the interactive dashboard \cite{chang2015package}.
\subsubsection*{$K$-means clustering} \label{kmean}     
The $K$-means clustering method is a popular unsupervised machine learning technique that is extensively utilised due to its simplicity and fast convergence. The $K$-means algorithm is a basic partitioning approach that utilises a distance metric for partitioning observations into clusters. The number of clusters, $K$, is determined beforehand. The centre of a cluster is known as the cluster centroid. Every data point is allocated to a cluster such that within a cluster the summed distance between the centroid and data points is minimised, and between clusters the summed distance between cluster centroids is maximised. Some distance metrics include Euclidean distance, Manhattan distance, cosine distance, Minkowski distance and correlation distance \cite{bora2014effect}. In this study, the Euclidean distance was adopted. The chosen value of $K$ directly influences both the convergence of the algorithm and the inferences. In this study, we considered a range of plausible value of $K$ and chose the value that gave the best fit, as determined by the silhouette coefficient, see section \ref{silh}. \\
The algorithm proceeds as follows. 1) Define the number of clusters $K$. 2) Randomly select $K$ data points as the cluster centroids. 3) Assign data points to the closest cluster centroid. 4) Recompute the cluster centroids. 5) Repeat steps 3) and 4) until either the centroids do not change or the maximum number of iterations is reached \cite{rathod2017design}. In this paper we apply the $K$-means algorithm to the proportion of vulnerable in a SA2 for each of the five AEDC domains and two indicators. \\
\subsubsection*{Cluster evaluation} \label{silh}
Internal and relative validation are two popular ways of evaluating a cluster analysis. Internal validation uses two fundamental principles to validate clusters: cohesion and separation. Cohesion measures average distance between items within clusters, while separation measures average distance of a cluster to the adjacent cluster. Clusters are confirmed in relative validation by altering the clustering algorithm's parameters, such as the number of clusters $K$, to optimise a given measure of fit.\\
In this study, we adopt the silhouette method for cluster evaluation \cite{kaufman2009finding}, which combines cohesion and separation. The similarity between the item and the cluster to which it belongs is represented by cohesion, and when compared to other clusters, it is described as separation. These comparisons may be quantified using the silhouette coefficient, which ranges from $-1$ to $1$, with a value near 1 suggesting good identification between the item and the cluster. In general, silhouette width scores less than 0.2 or silhouette width scores greater than 0.9 are problematic; silhouette width scores of 0.5 are good, and silhouette width scores between 0.7 and 0.9 are preferable \cite{rousseeuw1987silhouettes}. 
The Silhouette coefficient is given as:
\begin{equation}
    s(i)=\frac{b(i)-a(i)}{max\{a(i),b(i)\}}\\
    =\begin{cases} 
      1-\frac{a(i)}{b(i)} ,& a(i) < b(i) \\
      0 ,& a(i)=b(i) \\
      \frac{b(i)}{a(i)}-1, & a(i)> b(i),
   \end{cases}
\end{equation}
where $s(i)$ is the silhouette coefficient of data point $i$, $a(i)$ is the average distance between $i$ and all the other data points in the cluster to which $i$ belongs,  $a(i)$ represents the intra-cluster dissimilarity of sample $i$, $b(i)$ is the minimum average distance from $i$ to all clusters to which $i$ does not belong. The inter-cluster dissimilarity of sample $i$ is defined as $b(i)$.
\subsection*{R Shiny} \label{shiny}
Shiny is a R web application framework that allows the development of interactive web applications. This package makes it easy to create websites that interact with R without prior knowledge of web programming or other scripting languages. To create the shiny application, we uploaded the data set, built the clustering algorithm in R, and then used these two files to create the Shiny application. A brief summary
of Shiny and a description of the main components used to
implement the application are provided in Appendix \ref{shiny}. The program allows user involvement and generates interactive visualisations such as maps with padding and zooming capabilities.
One disadvantage of Shiny is that applications created with it can only be deployed online using the Shiny web server. It is noted, however, that although Shiny currently has a relatively limited feature set, this will likely expand, given the product's popularity \cite{rstudio2013shiny}.
It is important to note that the application requires access to the internet.
\section*{Results}  \label{res}
\begin{figure*}[htb!]
    \centering
    \includegraphics[scale=0.4]{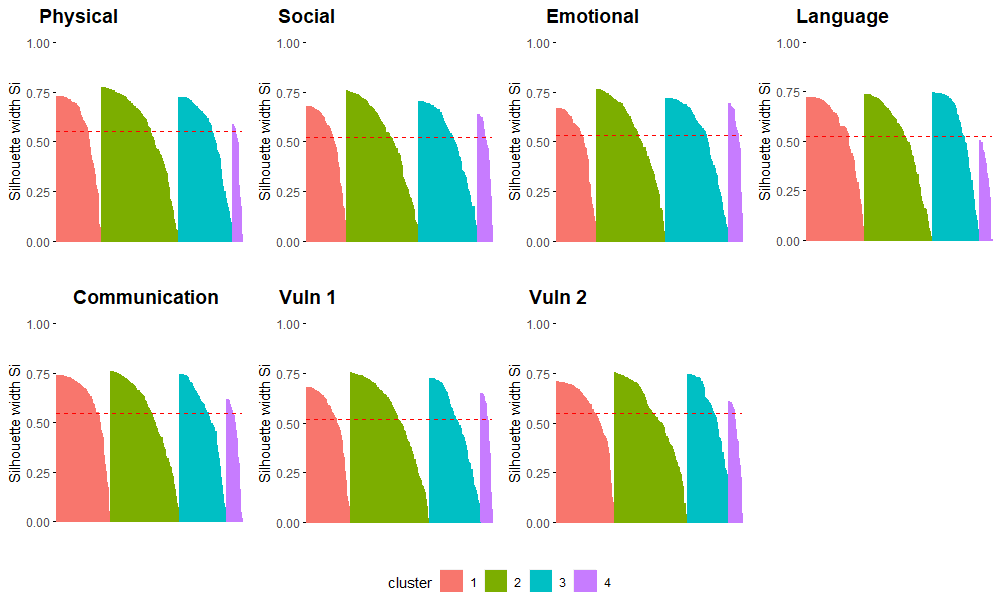}
    \caption{Silhouette plots for clusters across the five AEDC domains and two indicators, the $x$-axis are the clusters, and the height of each cluster is the silhouette width score for the cluster. The dotted line is the average silhouette width score across the four clusters.}
    \label{fig:my_label2}
\end{figure*}

\begin{figure*}[ht!]
    \centering
    \includegraphics[scale=0.4]{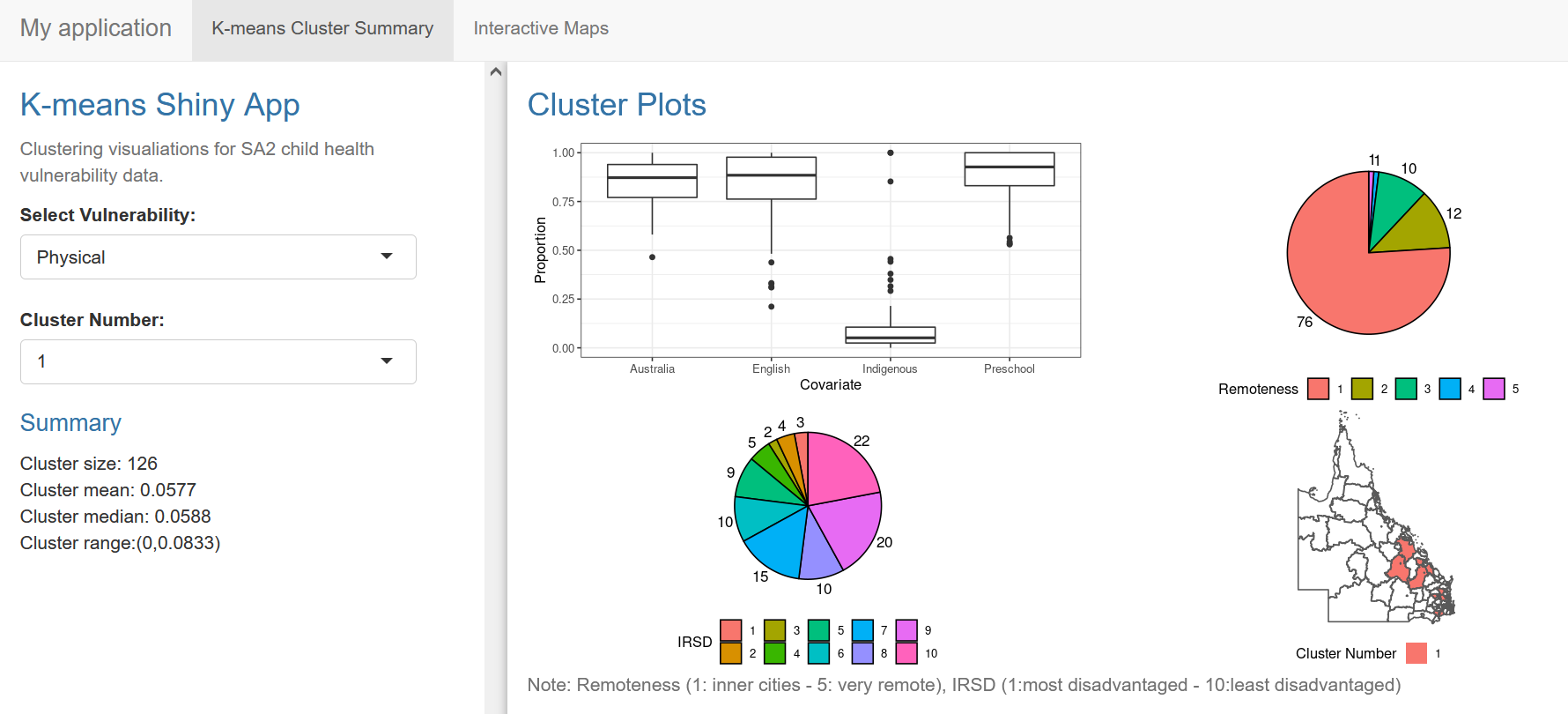}
    \caption{Example of $K$-means clustering results displayed in the web interface for C1 of the physical health development domain, the dashboard shows the box plot for the proportions of Australia, English, Indigenous and Preschool variables, and pie charts for the percentages of remoteness and IRSD and the location of C1 in Queensland map.}
    \label{fig:my_label3}
\end{figure*}

\begin{figure*}[ht!]
    \centering
    \includegraphics[scale=0.35]{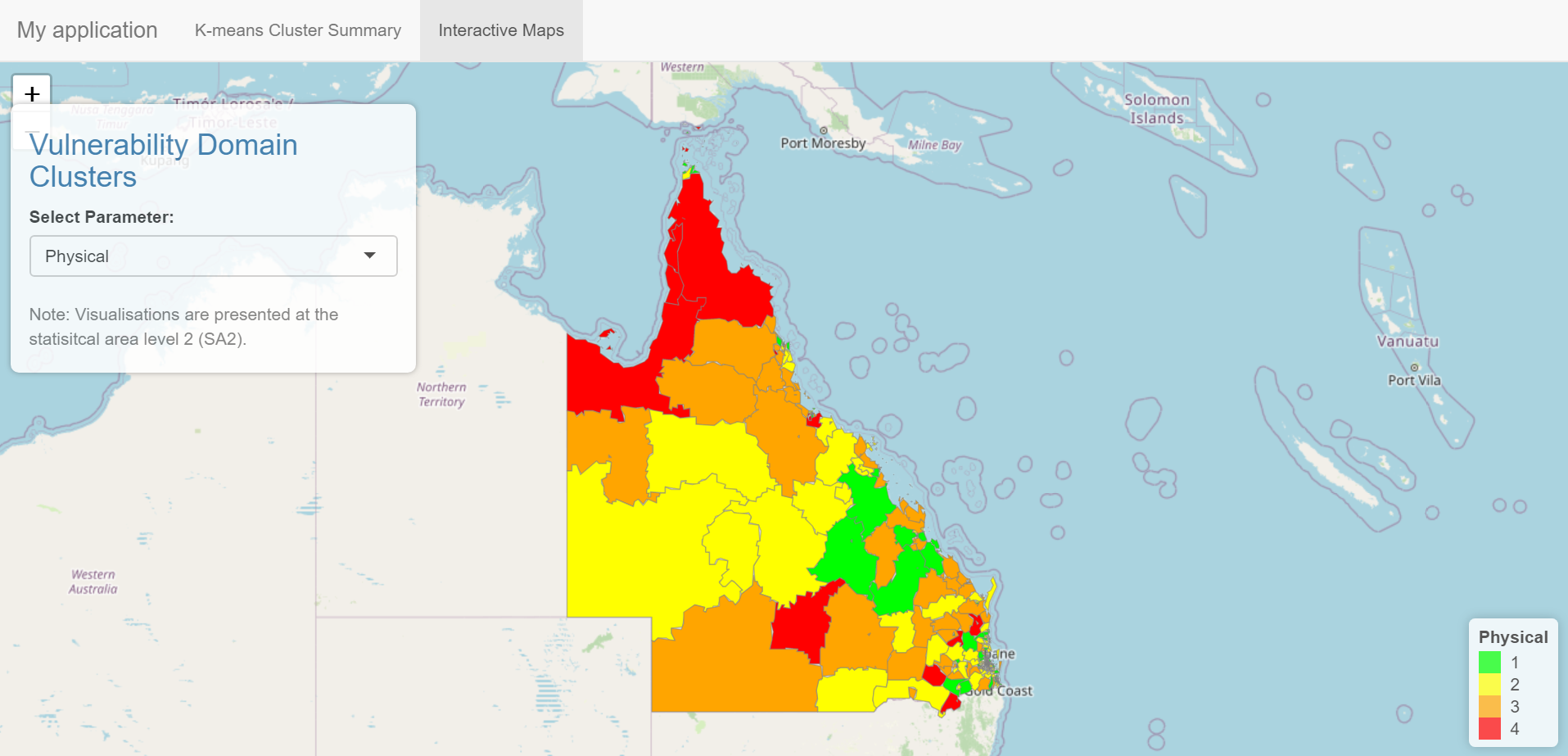}
    \caption{Map of the four clusters obtained for the physical health AEDC domain. The clusters are ordered from C1 (green, least vulnerable) to C4 (red, most vulnerable).}
    \label{fig:my_label4}
\end{figure*}

The developmentally vulnerable proportions were analysed on the log scale, due to skewness in the proportions, and converted back to their original scale in reporting the results. The number of clusters was evaluated for $K=\{3,4,5,...,12\}$, separately for each of the five AEDC domains and the two composite domain indicators (Vuln 1, Vuln 2). The optimal number of clusters for each domain was chosen to be four after validating the clusters internally using silhouette scores. The silhouette plots for the clusters in the five domains and two indicators are displayed in Figure \ref{fig:my_label2}. 
\\
Summary statistics for each cluster (size, mean, variance, range) for the five AEDC domains and two indicators with the association demographic factors are given in Appendix \ref{Appendix}. These results are visualised in the R Shiny application, accessed at
 \url{https://waladraidi.shinyapps.io/Shiny_2_6_2022/}.\\Figures \ref{fig:my_label3} and  \ref{fig:my_label4} illustrate the application interface. The interface includes two tabs.
For the first tab (Figure \ref{fig:my_label3}), which shows the $K$-means cluster summary, the user can select the type of development vulnerability and the cluster of interest. The clusters, labelled C1, C2, C3 and C4, correspond to vulnerability level ordered from lowest vulnerability (C1) to highest vulnerability (C4). Furthermore, this first tab shows the associated characteristics related to the demographic factors for each cluster and the location of the SA2 areas on the map. The second tab (Figure \ref{fig:my_label4}) shows a map of the distribution of the clusters (regions of differing vulnerability) for a given development vulnerability. The user can choose the type of AEDC domain from the five domains and two indicators, and can zoom in on the Queensland map to view finer details for each region. This provides an interactive visual summary of vulnerability across the regions of Queensland, and comparison of the vulnerabilities across the five AEDC domains and two indicators. An example map is given in Figure \ref{fig:my_label4}. A selection of outputs from the app, comparing C4 (highest vulnerability) to C1 (lowest vulnerability), is provided in Appendix \ref{Shiny}. \\
\begin{table*}[h!]
\caption{Comparison of clusters, C4 (most vulnerable) and C1 (least vulnerable), averaged over the five AEDC domains (excluding Vuln 1 and  Vuln 2). The percentage of children where 'English Not Primary Language' was calculated as 1 minus 'English'. Similarly, 'No Preschool' is 1 minus 'Preschool')}
\begin{tabular}{|l|ll|ll|}
\hline
                      & \multicolumn{2}{c}{C4(\%)} & \multicolumn{2}{c|}{C1(\%)} \\ \hline
                      & mean        & sd          & mean         & sd          \\
Vulnerable            & 24.6         & 63.9        & 4.60          & 1.10         \\
\begin{tabular}[c]{@{}l@{}}English Not Primary Language\end{tabular} &
  19.0 &
  12.7 &
  14.4 &
  1.50 \\
Indigenous            & 39.6        & 17.1        & 9.20          & 1.60         \\
No Preschool          & 23.2        & 1.80         & 11.8         & 0.80        \\
Remoteness – Cities &
   31.8&
   10.6 &
   70.2 &
  5.90 \\
Remoteness – Regional & 49.2         & 6.40         & 28.0         & 5.10         \\
Remoteness – Remote   & 11.4         & 5.60         & 2.40          & 0.90         \\
IRSD – Low            & 84.4         & 13.2        & 16.2         & 6.20         \\ 

\hline
\end{tabular}
\label{tab:my-tableco}
\end{table*}
Averaging over the five AEDC domains, the SA2s that make up C4 have an average of 25\% of children identified as vulnerable, compared to 5\% for C1 (Table \ref{tab:my-tableco}); this discrepancy is also observed within each AEDC domain (Table \ref{tab:my-table10}). In comparing the most vulnerable cluster, C4, to the least vulnerable cluster, C1, there are higher proportions of children who do not have English as their mother language, who are Indigenous and who did not attend preschool (Tables \ref{tab:my-tableco} and \ref{tab:my-table10} ). The one exception to this is that, for the SA2s making up the most Emotionally vulnerable cluster, C4, there is a higher proportion of children who do have English as their mother language compared to the least Emotionally vulnerable cluster, C1 (Table \ref{tab:my-table10}). The SA2s belonging to C4 are located in far north and north west Queensland, and a small number can be found in the coastal areas of Queensland (Table \ref{tab:my-table100}). In contrast, the SA2s belonging to C1 can be found in the south east of the state. This region contains the majority of the children of Queensland and the capital city, Brisbane (Table \ref{tab:my-table100}). Across all AEDC domains, there is a much higher proportion of children residing in SA2s belonging to C4 with a low IRSD score (greater socio-economic disadvantage) compared to C1.\\
In comparing Vuln 1 and Vuln 2, unsurprisingly, the proportion of children who are vulnerable on two or more (2+) domains is lower than the proportion vulnerable on one or more (1+) domains for both C4 and C1. For C4, the proportion of children who don't have English as their primary language is higher for 2+ vulnerabilities compared to 1+ vulnerabilities. In comparison, the proportion of children who identified as Indigenous is lower for 2+ vulnerabilities compared to 1+ vulnerabilities. The proportion who did not attend preschool is about the same for 1+ and 2+ vulnerabilities. The SA2s belonging to C4 for Vuln 1 are located in the same geographic areas of Queensland as Vuln 2 and additionally in the southeast and central coast. For the SA2s in C4, there is a higher proportion of children residing in SA2s with a low IRSD score (greater socio-economic disadvantage) for Vuln 1  compared to Vuln 2. \\
 In comparison, across the five domains for the most vulnerable cluster (C4), the smallest cluster size can be found in the physical health domain with around 30 SA2 areas, and the largest cluster size can be found in the communication skills domain, 46 SA2 areas. In addition, there was a notably higher proportion of Indigenous children in Physical domains in comparison with the rest of the AEDC domains, the proportion of the country of birth was greater than 85\% across all the clusters, and for all domains, with some slight differences between the clusters of no more than around 5\%. 
\section*{Discussion}
In this study, the $K$-means algorithm was applied to investigate commonalities  in statistical areas across Queensland, Australia, with respect to children's vulnerability based on five AEDC domains and two indicators. Four clusters were identified for each of these domains, and demographic profiles were developed for each cluster. In addition to presenting summary statistics in tabular form, an R Shiny app was developed to visualise and summarise the results of the analyses. This app enables users to engage with the results interactively. For example, health managers can use the app to identify regions with high proportions of developmentally vulnerable children and develop more targeted services for these areas. This study is crucial for the government and individuals to identify the regions of high vulnerabilities and improve services for these population groups. \\
 The clustering analyses reveal a strong relationship between AEDC domains and socio-economic and remoteness factors. We found that SA2s with the lowest proportion of vulnerable children typically had larger proportions of children who attended pre-school and whose primary language is English. However, there was substantial spatial variation in the results.\\
The communication skills domain (Communication) was found to have the largest cluster size for the most vulnerable SA2s (C4) compared to the other domains. In contrast, the language and cognitive skill domain (Language) had the largest cluster size for the least vulnerable SA2s (C1) compared to other domains. SA2s in this later group were characterised by children typically from high SA2 socio-economic regions with a lower proportion of Indigenous status and a higher proportion of attendance at pre-school.In this case study, the data are analysed at the SA2 level of aggregation. Therefore, care must be taken in making inferences at another level of aggregation or about individuals due to biases such as  Simpson's paradox \cite{heydtmann2002nature}. \\
The clustering of these SA2 level child developmentally vulnerable data offers a comprehensive breakdown of the factors impacting child health development across Queensland. This breakdown of vulnerabilities at the statistical area level allows for improved region-based analysis and policy development.
\begin{appendix}

\end{appendix}


\begin{backmatter}

\section*{Acknowledgements}
The authors thank the Children's  Health Queensland staff for their support during the research. 

\section*{Funding}
This research was supported by Children's Health Queensland (CHQ) and The Queensland University of Technology - Center for Data Science, Queensland, Australia.
\section*{Availability of data and materials}
All the data used in this study are available to the public from the Australian Bureau of Statistic and Australian Early Development Census.
\section*{Declaration}
“This paper uses data from the Australian Early Development Census (AEDC). The AEDC is funded by the Australian Government Department of Education, Skills and Employment. The findings and views reported are those of the author and should not be attributed to the Department or the Australian Government.”
\section*{Competing interests}
All authors have read and approved this version of the article, and declared that they have no competing financial or non-financial interests to disclose.

\section*{Consent for publication}
All the authors have provided consent to publish this manuscript

\section*{Authors' contributions}
Study design and interpretation of analyses: Areed, Price, Thompson, Mengersen and Arnett. Data analysis and manuscript drafting: Areed.
Critical revision of the manuscript: Thompson,Mengersen, Areed, Price, and Arnett. 

\bibliographystyle{bmc-mathphys} 
\bibliography{bmc_article}      

\end{backmatter}

\begin{appendix}
\onecolumn

\counterwithin{figure}{section}
\counterwithin{table}{section}
\renewcommand\thefigure{\thesection\arabic{figure}}
\renewcommand\thetable{\thesection\arabic{table}}
\section{Results from $K$-means algorithm for each type of developmentally vulnerable. For all results, the clusters are ordered from lowest vulnerability (C1) to highest vulnerability (C4).}  \label{Appendix}

\begin{table*}[!htbp]
\centering
\caption{$K$-means results for physical health domain vulnerability (Physical), where $n$ is the number of SA2's in the cluster.}
\scalebox{0.6}{
\begin{tabular}{llllllllllll}
\hline
                     & C1 (n=126) &              &  & C2 (n=219) &               &  & C3 (n=151) &              &  & C4 (n=30) &              \\ \cline{2-3} \cline{5-6} \cline{8-9} \cline{11-12} 
                     & mean       & range        &  & mean       & range         &  & mean       & range        &  & mean      & range        \\ \cline{2-3} \cline{5-6} \cline{8-9} \cline{11-12} 
\textbf{Domain}      &            &              &  &            &               &  &            &              &  &           &              \\
Physical             & 0.06       & (0.00, 0.08) &  & 0.11       & (0.08, 0.14)  &  & 0.17       & (0.14, 0.22) &  & 0.28      & (0.22, 0.68) \\
\textbf{Demographic} &            &              &  &            &               &  &            &              &  &           &              \\
Australia            & 0.85       & (0.46, 1.00) &  & 0.88       & (0.41, 1.00)  &  & 0.89       & (0.52, 1.00) &  & 0.89      & (0.72, 1.00) \\
English              & 0.85       & (0.21, 1.00) &  & 0.81       & (0.06, 1.00)  &  & 0.83       & (0.34, 1.00) &  & 0.69      & (0.01, 1.00) \\
Indigenous           & 0.10       & (0.00, 1.00) &  & 0.13       & (0.00, 1.00)  &  & 0.21       & (0.00, 0.83) &  & 0.40      & (0.00, 1.00) \\
Preschool            & 0.89       & (0.53, 1.00) &  & 0.85       & (0.44, 1.00 ) &  & 0.78       & (0.44, 1.00) &  & 0.7579    & (0.46, 1.00) \\
Remoteness           &            &              &  &            &               &  &            &              &  &           &              \\
Inner cities         & 0.76       &              &  & 0.56       &               &  & 0.43       &              &  & 0.23      &              \\
Inner regional       & 0.12       &              &  & 0.22       &               &  & 0.28       &              &  & 0.27      &              \\
Outer regional       & 0.10       &              &  & 0.18       &               &  & 0.24       &              &  & 0.30      &              \\
remote               & 0.01       &              &  & 0.02       &               &  & 0.03       &              &  & 0.03      &              \\
Very remote          & 0.01       &              &  & 0.02       &               &  & 0.02       &              &  & 0.17      &              \\
IRSD                 &            &              &  &            &               &  &            &              &  &           &              \\
1 (low)              & 0.03       &              &  & 0.05       &               &  & 0.15       &              &  & 0.50      &              \\
2                    & 0.04       &              &  & 0.07       &               &  & 0.18       &              &  & 0.14      &              \\
3                    & 0.02       &              &  & 0.08       &               &  & 0.16       &              &  & 0.30      &              \\
4                    & 0.05       &              &  & 0.12       &               &  & 0.12       &              &  & 0.00      &              \\
5                    & 0.09       &              &  & 0.11       &               &  & 0.10       &              &  & 0.00      &              \\
6                    & 0.10       &              &  & 0.17       &               &  & 0.11       &              &  & 0.00      &              \\
7                    & 0.15       &              &  & 0.10       &               &  & 0.05       &              &  & 0.03      &              \\
8                    & 0.10       &              &  & 0.13       &               &  & 0.07       &              &  & 0.00      &              \\
9                    & 0.20       &              &  & 0.09       &               &  & 0.03       &              &  & 0.03      &              \\
10 (high)            & 0.22       &              &  & 0.08       &               &  & 0.03       &              &  & 0.00      &              \\ \hline
\end{tabular}}

\label{tab:my-tableW1}
\end{table*}

\begin{table*}[!htbp]
\centering
\caption{$K$-means results for social competence  domain vulnerability (Social), where $n$ is the number of SA2's in the cluster.}
\scalebox{0.6}{
\begin{tabular}{llllllllllll}
\hline
                     & C1 (n=111) &              &  & C2 (n=205) &              &  & C3 (n=168) &              &  & C4 (n=42) &              \\ \cline{2-3} \cline{5-6}  \cline{8-9} \cline{11-12} 
                     & mean       & range        &  & mean       & range        &  & mean       & range        &  & mean      & range        \\ \cline{2-3} \cline{5-6} \cline{8-9} \cline{11-12} 
\textbf{Domain}      &            &              &  &            &              &  &            &              &  &           &              \\
Social               & 0.04       & (0.00, 0.07) &  & 0.097      & (0.07, 0.12) &  & 0.15       & (0.12, 0.19) &  & 0.23      & (0.19, 0.39) \\
\textbf{Demographic} &            &              &  &            &              &  &            &              &  &           &              \\
Australia            & 0.85       & (0.41, 1.00) &  & 0.88       & (0.53, 1.00) &  & 0.89       & (0.46, 1.00) &  & 0.89      & (0.53, 1.00) \\
English              & 0.84       & (0.14, 1.00) &  & 0.82       & (0.21, 1.00) &  & 0.82       & (0.19, 1.00) &  & 0.75      & (0.01, 1.00) \\
Indigenous           & 0.10       & (0.00, 0.85) &  & 0.15       & (0.00, 1.00) &  & 0.19       & (0.00, 1.00) &  & 0.28      & (0.00, 1.00) \\
Preschool            & 0.88       & (0.44, 1.00) &  & 0.84       & (0.45, 1.00) &  & 0.81       & (0.44, 1.00) &  & 0.78      & (0.46, 1.00) \\
Remoteness           &            &              &  &            &              &  &            &              &  &           &              \\
Inner cities         & 0.66       &              &  & 0.57       &              &  & 0.51       &              &  & 0.45      &              \\
Inner regional       & 0.20       &              &  & 0.19       &              &  & 0.26       &              &  & 0.19      &              \\
Outer regional       & 0.11       &              &  & 0.20       &              &  & 0.18       &              &  & 0.29      &              \\
remote               & 0.02       &              &  & 0.01       &              &  & 0.03       &              &  & 0.00      &              \\
Very remote          & 0.01       &              &  & 0.03       &              &  & 0.02       &              &  & 0.07      &              \\
IRSD                 &            &              &  &            &              &  &            &              &  &           &              \\
1 (low)              & 0.02       &              &  & 0.07       &              &  & 0.13       &              &  & 0.31      &              \\
2                    & 0.04       &              &  & 0.05       &              &  & 0.16       &              &  & 0.24      &              \\
3                    & 0.04       &              &  & 0.09       &              &  & 0.14       &              &  & 0.12      &              \\
4                    & 0.08       &              &  & 0.13       &              &  & 0.08       &              &  & 0.04      &              \\
5                    & 0.09       &              &  & 0.12       &              &  & 0.08       &              &  & 0.05      &              \\
6                    & 0.09       &              &  & 0.14       &              &  & 0.15       &              &  & 0.07      &              \\
7                    & 0.14       &              &  & 0.09       &              &  & 0.08       &              &  & 0.05      &              \\
8                    & 0.11       &              &  & 0.12       &              &  & 0.08       &              &  & 0.05      &              \\
9                    & 0.21       &              &  & 0.10       &              &  & 0.04       &              &  & 0.02      &              \\
10 (high)            & 0.18       &              &  & 0.09       &              &  & 0.06       &              &  & 0.05      &              \\ \hline
\end{tabular}}
\label{tab:my-tableW2}
\end{table*}

\begin{table*}[!htbp]
\centering
\caption{$K$-means results for emotional maturity domain vulnerability (Emotional), where $n$ is the number of SA2's in the cluster.}
\scalebox{0.6}{
\begin{tabular}{llllllllllll}
\hline
                     & C1 (n=113) &              &  & C2 (n=194) &              &  & C3 (n=180) &              &  & C4 (n=39) &              \\ \cline{2-3} \cline{5-6} \cline{8-9} \cline{11-12} 
                     & mean       & range        &  & mean       & range        &  & mean       & range        &  & mean      & range        \\ \cline{2-3} \cline{5-6} \cline{8-9} \cline{11-12} 
\textbf{Domain}      &            &              &  &            &              &  &            &              &  &           &              \\
Emotional            & 0.05       & (0.00, 0.07) &  & 0.09       & (0.07, 0.10) &  & 0.13       & (0.11, 0.16) &  & 0.20      & (0.17, 0.27) \\
\textbf{Demographic} &            &              &  &            &              &  &            &              &  &           &              \\
Australia            & 0.87       & (0.41, 1.00) &  & 0.87       & (0.46, 1.00) &  & 0.89       & (0.53, 1.00) &  & 0.9034    & (0.70, 1.00) \\
English              & 0.86       & (0.21, 1.00) &  & 0.82       & (0.27, 1.00) &  & 0.81       & (0.14, 1.00) &  & 0.77      & (0.01, 1.00) \\
Indigenous           & 0.11       & (0.00, 0.85) &  & 0.14       & (0.00, 1.00) &  & 0.19       & (0.00, 1.00) &  & 0.28      & (0.00, 1.00) \\
Preschool            & 0.87       & (0.44, 1.00) &  & 0.84       & (0.45, 1.00) &  & 0.82       & (0.44, 1.00) &  & 0.78      & (0.45, 0.99) \\
Remoteness           &            &              &  &            &              &  &            &              &  &           &              \\
Inner cities         & 0.64       &              &  & 0.56       &              &  & 0.51       &              &  & 0.51      &              \\
Inner regional       & 0.18       &              &  & 0.22       &              &  & 0.24       &              &  & 0.20      &              \\
Outer regional       & 0.15       &              &  & 0.19       &              &  & 0.19       &              &  & 0.21      &              \\
remote               & 0.02       &              &  & 0.01       &              &  & 0.03       &              &  & 0.00      &              \\
Very remote          & 0.01       &              &  & 0.02       &              &  & 0.03       &              &  & 0.08      &              \\
IRSD                 &            &              &  &            &              &  &            &              &  &           &              \\
1 (low)              & 0.03       &              &  & 0.05       &              &  & 0.13       &              &  & 0.38      &              \\
2                    & 0.02       &              &  & 0.08       &              &  & 0.17       &              &  & 0.10      &              \\
3                    & 0.07       &              &  & 0.08       &              &  & 0.12       &              &  & 0.18      &              \\
4                    & 0.14       &              &  & 0.10       &              &  & 0.07       &              &  & 0.05      &              \\
5                    & 0.10       &              &  & 0.09       &              &  & 0.13       &              &  & 0.00      &              \\
6                    & 0.10       &              &  & 0.16       &              &  & 0.12       &              &  & 0.08      &              \\
7                    & 0.09       &              &  & 0.13       &              &  & 0.06       &              &  & 0.10      &              \\
8                    & 0.11       &              &  & 0.12       &              &  & 0.08       &              &  & 0.03      &              \\
9                    & 0.16       &              &  & 0.10       &              &  & 0.06       &              &  & 0.05      &              \\
10 (high)            & 0.18       &              &  & 0.09       &              &  & 0.06       &              &  & 0.03      &              \\ \hline
\end{tabular}}

\label{tab:my-tableW3}
\end{table*}

\begin{table*}[!htbp]
\centering
\caption{$K$-means results for language domain vulnerability (Language), where $n$ is the number of SA2's in the cluster.}
\scalebox{0.6}{
\begin{tabular}{llllllllllll}
\hline
                     & C1 (n=162) &              &  & C2 (n=194) &              &  & C3 (n=133) &              &  & C4 (n=37) &              \\ \cline{2-3} \cline{5-6} \cline{8-9} \cline{11-12} 
                     & mean       & range        &  & mean       & range        &  & mean       & range        &  & mean      & range        \\ \cline{2-3} \cline{5-6} \cline{8-9} \cline{11-12} 
\textbf{Domain}      &            &              &  &            &              &  &            &              &  &           &              \\
Language             & 0.03       & (0.00, 0.05) &  & 0.08       & (0.06, 0.11) &  & 0.14       & (0.11, 0.21) &  & 0.29      & (0.22, 0.55) \\
\textbf{Demographic} &            &              &  &            &              &  &            &              &  &           &              \\
Australia            & 0.85       & (0.46, 1.00) &  & 0.89       & (0.41, 1.00) &  & 0.90       & (0.59, 1.00) &  & 0.89      & (0.73, 1.00) \\
English              & 0.85       & (0.14, 1.00) &  & 0.82       & (0.21, 1.00) &  & 0.79       & (0.01, 1.00) &  & 0.56      & (0.06, 0.83) \\
Indigenous           & 0.07       & (0.00, 0.64) &  & 0.16       & (0.00, 0.85) &  & 0.27       & (0.00, 1.00) &  & 0.69      & (0.15, 1.00) \\
Preschool            & 0.89       & (0.45, 1.00) &  & 0.82       & (0.44, 1.00) &  & 0.76       & (0.45, 1.00) &  & 0.78      & (0.46, 0.97) \\
Remoteness           &            &              &  &            &              &  &            &              &  &           &              \\
Inner cities         & 0.77       &              &  & 0.52       &              &  & 0.32       &              &  & 0.00      &              \\
Inner regional       & 0.12       &              &  & 0.26       &              &  & 0.29       &              &  & 0.18      &              \\
Outer regional       & 0.11       &              &  & 0.19       &              &  & 0.27       &              &  & 0.36      &              \\
remote               & 0.00       &              &  & 0.03       &              &  & 0.03       &              &  & 0.09      &              \\
Very remote          & 0.00       &              &  & 0.00       &              &  & 0.08       &              &  & 0.36      &              \\
IRSD                 &            &              &  &            &              &  &            &              &  &           &              \\
1 (low)              & 0.01       &              &  & 0.07       &              &  & 0.21       &              &  & 0.82      &              \\
2                    & 0.02       &              &  & 0.09       &              &  & 0.23       &              &  & 0.09      &              \\
3                    & 0.02       &              &  & 0.11       &              &  & 0.21       &              &  & 0.09      &              \\
4                    & 0.05       &              &  & 0.12       &              &  & 0.13       &              &  & 0.00      &              \\
5                    & 0.08       &              &  & 0.13       &              &  & 0.08       &              &  & 0.00      &              \\
6                    & 0.11       &              &  & 0.19       &              &  & 0.07       &              &  & 0.00      &              \\
7                    & 0.15       &              &  & 0.10       &              &  & 0.01       &              &  & 0.00      &              \\
8                    & 0.13       &              &  & 0.11       &              &  & 0.03       &              &  & 0.00      &              \\
9                    & 0.19       &              &  & 0.06       &              &  & 0.02       &              &  & 0.00      &              \\
10 (high)            & 0.22       &              &  & 0.03       &              &  & 0.01       &              &  & 0.00      &              \\ \hline
\end{tabular}}

\label{tab:my-tableW4}
\end{table*}

\begin{table*}[!htbp]
\centering
\caption{$K$-means results for communication skills domain vulnerability (Communication), where $n$ is the number of SA2's in the cluster.}
\scalebox{0.6}{
\begin{tabular}{llllllllllll}
\hline
                     & C1 (n=152) &              &  & C2 (n=195) &              &  & C3 (n=133) &              &  & C4 (n=46) &               \\ \cline{2-3} \cline{5-6} \cline{8-9} \cline{11-12} 
                     & mean       & range        &  & mean       & range        &  & mean       & range        &  & mean      & range         \\ \cline{2-3} \cline{5-6} \cline{8-9} \cline{11-12} 
\textbf{Domain}      &            &              &  &            &              &  &            &              &  &           &               \\
Communication        & 0.04       & (0.00, 0.07) &  & 0.09       & (0.07, 0.12) &  & 0.15       & (0.12, 0.18) &  & 0.22      & (0.18, 0.43)  \\
\textbf{Demographic} &            &              &  &            &              &  &            &              &  &           &               \\
Australia            & 0.88       & (0.41, 1.00) &  & 0.86       & (0.59, 1.00) &  & 0.89       & (0.63, 1.00) &  & 0.88      & (0.67, 1.00 ) \\
English              & 0.88       & (0.58, 1.00) &  & 0.84       & (0.21, 1.00) &  & 0.77       & (0.14, 1.00) &  & 0.65      & (0.01, 1.00)  \\
Indigenous           & 0.08       & (0.00, 0.46) &  & 0.14       & (0.00, 0.85) &  & 0.24       & (0.00, 1.00) &  & 0.33      & (0.00, 1.00)  \\
Preschool            & 0.88       & (0.44, 1.00) &  & 0.85       & (0.46, 1.00) &  & 0.78       & (0.45, 1.00) &  & 0.74      & (0.44, 1.00)  \\
Remoteness           &            &              &  &            &              &  &            &              &  &           &               \\
Inner cities         & 0.68       &              &  & 0.58       &              &  & 0.43       &              &  & 0.40      &               \\
Inner regional       & 0.18       &              &  & 0.21       &              &  & 0.26       &              &  & 0.19      &               \\
Outer regional       & 0.13       &              &  & 0.18       &              &  & 0.21       &              &  & 0.27      &               \\
remote               & 0.01       &              &  & 0.02       &              &  & 0.03       &              &  & 0.03      &               \\
Very remote          & 0.00       &              &  & 0.01       &              &  & 0.07       &              &  & 0.11      &               \\
IRSD                 &            &              &  &            &              &  &            &              &  &           &               \\
1 (low)              & 0.01       &              &  & 0.06       &              &  & 0.17       &              &  & 0.38      &               \\
2                    & 0.03       &              &  & 0.07       &              &  & 0.19       &              &  & 0.24      &               \\
3                    & 0.03       &              &  & 0.08       &              &  & 0.19       &              &  & 0.13      &               \\
4                    & 0.06       &              &  & 0.11       &              &  & 0.12       &              &  & 0.11      &               \\
5                    & 0.07       &              &  & 0.12       &              &  & 0.12       &              &  & 0.02      &               \\
6                    & 0.15       &              &  & 0.17       &              &  & 0.07       &              &  & 0.03      &               \\
7                    & 0.12       &              &  & 0.13       &              &  & 0.05       &              &  & 0.00      &               \\
8                    & 0.15       &              &  & 0.09       &              &  & 0.06       &              &  & 0.03      &               \\
9                    & 0.16       &              &  & 0.11       &              &  & 0.01       &              &  & 0.03      &               \\
10 (high)            & 0.22       &              &  & 0.06       &              &  & 0.02       &              &  & 0.03      &               \\ \hline
\end{tabular}}

\label{tab:my-tableW5}
\end{table*}

\begin{table*}[!htbp]
\centering
\caption{$K$-means results for vulnerability on one or more domain(s) (Vuln 1), where $n$ is the number of SA2's in the cluster.}
\scalebox{0.6}{
\begin{tabular}{llllllllllll}
\hline
                     & C1 (n=101) &              &  & C2 (n=181) &              &  & C3 (n=173) &              &  & C4 (n=71) &              \\ \cline{2-3} \cline{5-6} \cline{8-9} \cline{11-12} 
                     & mean       & range        &  & mean       & range        &  & mean       & range        &  & mean      & range        \\ \cline{2-3} \cline{5-6} \cline{8-9} \cline{11-12} 
\textbf{Domain}      &            &              &  &            &              &  &            &              &  &           &              \\
Vuln 1               & 0.15       & (0.06, 0.19) &  & 0.24       & (0.19, 0.28) &  & 0.29       & (0.28, 0.38) &  & 0.40      & (0.38, 0.71) \\
\textbf{Demographic} &            &              &  &            &              &  &            &              &  &           &              \\
Australia            & 0.85       & (0.46, 1.00) &  & 0.87       & (0.41, 1.00) &  & 0.89       & (0.53, 1.00) &  & 0.90      & (0.66, 1.00) \\
English              & 0.88       & (0.46, 1.00) &  & 0.84       & (0.29, 1.00) &  & 0.81       & (0.14, 1.00) &  & 0.6993    & (0.01, 1.00) \\
Indigenous           & 0.07       & (0.00, 0.38) &  & 0.13       & (0.00, 1.00) &  & 0.19       & (0.00, 0.93) &  & 0.33      & (0.00, 1.00) \\
Preschool            & 0.90       & (0.44, 1.00) &  & 0.85       & (0.54, 1.00) &  & 0.80       & (0.44, 1.00) &  & 0.79      & (0.45, 1.00) \\
Remoteness           &            &              &  &            &              &  &            &              &  &           &              \\
Inner cities         & 0.75       &              &  & 0.56       &              &  & 0.46       &              &  & 0.35      &              \\
Inner regional       & 0.15       &              &  & 0.22       &              &  & 0.26       &              &  & 0.23      &              \\
Outer regional       & 0.09       &              &  & 0.18       &              &  & 0.22       &              &  & 0.21      &              \\
remote               & 0.01       &              &  & 0.02       &              &  & 0.03       &              &  & 0.03      &              \\
Very remote          & 0.00       &              &  & 0.02       &              &  & 0.03       &              &  & 0.18      &              \\
IRSD                 &            &              &  &            &              &  &            &              &  &           &              \\
1 (low)              & 0.00       &              &  & 0.03       &              &  & 0.14       &              &  & 0.50      &              \\
2                    & 0.02       &              &  & 0.04       &              &  & 0.22       &              &  & 0.12      &              \\
3                    & 0.05       &              &  & 0.06       &              &  & 0.17       &              &  & 0.26      &              \\
4                    & 0.06       &              &  & 0.11       &              &  & 0.10       &              &  & 0.09      &              \\
5                    & 0.08       &              &  & 0.13       &              &  & 0.12       &              &  & 0.00      &              \\
6                    & 0.11       &              &  & 0.13       &              &  & 0.11       &              &  & 0.03      &              \\
7                    & 0.10       &              &  & 0.17       &              &  & 0.05       &              &  & 0.00      &              \\
8                    & 0.14       &              &  & 0.16       &              &  & 0.03       &              &  & 0.00      &              \\
9                    & 0.20       &              &  & 0.11       &              &  & 0.02       &              &  & 0.00      &              \\
10 (high)            & 0.24       &              &  & 0.10       &              &  & 0.04       &              &  & 0.00      &              \\ \hline
\end{tabular}}

\label{tab:my-table}
\end{table*}

\begin{table*}[!htbp]
\centering
\caption{$K$-means results for vulnerability on two or more domains (Vuln 2), where $n$ is the number of SA2's in the cluster.}
\scalebox{0.6}{
\begin{tabular}{llllllllllll}
\hline
                     & C1 (n=162) &              &  & C2 (n=207) &              &  & C3 (n=117) &              &  & C4 (n=40) &              \\ \cline{2-3} \cline{5-6} \cline{8-9} \cline{11-12} 
                     & mean       & range        &  & mean       & range        &  & mean       & range        &  & mean      & range        \\ \cline{2-3} \cline{5-6} \cline{8-9} \cline{11-12} 
\textbf{Domain}      &            &              &  &            &              &  &            &              &  &           &              \\
Vuln 2               & 0.07       & (0.00, 0.10) &  & 0.13       & (0.10, 0.16) &  & 0.19       & (0.16, 0.23) &  & 0.28      & (0.24, 0.55) \\
\textbf{Demographic} &            &              &  &            &              &  &            &              &  &           &              \\
Australia            & 0.86       & (0.46, 1.00) &  & 0.88       & (0.41, 1.00) &  & 0.89       & (0.53, 1.00) &  & 0.90      & (0.72, 1.00) \\
English              & 0.85       & (0.14, 1.00) &  & 0.83       & (0.14, 1.00) &  & 0.81       & (0.31, 100)  &  & 0.66      & (0.01, 1.00) \\
Indigenous           & 0.13       & (0.00, 1.00) &  & 0.33       & (0.00, 1.00) &  & 0.07       & (0.00, 0.38) &  & 0.19      & (0.00, 0.93) \\
Preschool            & 0.85       & (0.54, 1.00) &  & 0.79       & (0.45, 1.00) &  & 0.90       & (0.44, 1.00) &  & 0.80      & (0.44, 1.00) \\
Remoteness           &            &              &  &            &              &  &            &              &  &           &              \\
Inner cities         & 0.70       &              &  & 0.57       &              &  & 0.42       &              &  & 0.35      &              \\
Inner regional       & 0.17       &              &  & 0.20       &              &  & 0.29       &              &  & 0.25      &              \\
Outer regional       & 0.13       &              &  & 0.18       &              &  & 0.22       &              &  & 0.25      &              \\
remote               & 0.00       &              &  & 0.02       &              &  & 0.03       &              &  & 0.02      &              \\
Very remote          & 0.00       &              &  & 0.03       &              &  & 0.04       &              &  & 0.13      &              \\
IRSD                 &            &              &  &            &              &  &            &              &  &           &              \\
1 (low)              & 0.02       &              &  & 0.04       &              &  & 0.15       &              &  & 0.52      &              \\
2                    & 0.03       &              &  & 0.06       &              &  & 0.22       &              &  & 0.20      &              \\
3                    & 0.04       &              &  & 0.07       &              &  & 0.19       &              &  & 0.20      &              \\
4                    & 0.07       &              &  & 0.13       &              &  & 0.10       &              &  & 0.02      &              \\
5                    & 0.09       &              &  & 0.12       &              &  & 0.11       &              &  & 0.00      &              \\
6                    & 0.10       &              &  & 0.20       &              &  & 0.08       &              &  & 0.0       &              \\
7                    & 0.13       &              &  & 0.12       &              &  & 0.04       &              &  & 0.03      &              \\
8                    & 0.12       &              &  & 0.13       &              &  & 0.05       &              &  & 0.00      &              \\
9                    & 0.20       &              &  & 0.07       &              &  & 0.02       &              &  & 0.03      &              \\
10 (high)            & 0.20       &              &  & 0.06       &              &  & 0.04       &              &  & 0.00      &              \\ \hline
\end{tabular}}

\label{tab:my-tableW7}
\end{table*}
\noindent

\begin{sidewaystable*}[!htbp]
\caption{Comparison of C4 (most vulnerable) to C1 (least vulnerable) for the five domains of development and two indicators.}
\scalebox{1}{
\begin{tabular}{|l|ll|ll|ll|ll|ll|ll|ll|}
\hline
 &
  \multicolumn{2}{c|}{Physical} &
  \multicolumn{2}{c|}{Social} &
  \multicolumn{2}{c|}{Emotional} &
  \multicolumn{2}{c|}{Language} &
  \multicolumn{2}{c|}{Communication} &
  \multicolumn{2}{c|}{Vuln 1} &
  \multicolumn{2}{c|}{Vuln 2} \\ \hline
Cluster size &
  \begin{tabular}[c]{@{}l@{}}C4(\%)\\ (n-30)\end{tabular} &
  \begin{tabular}[c]{@{}l@{}}C1(\%)\\ (n=126)\end{tabular} &
  \begin{tabular}[c]{@{}l@{}}C4(\%)\\ (n=42)\end{tabular} &
  \begin{tabular}[c]{@{}l@{}}C1(\%)\\ (n=111)\end{tabular} &
  \begin{tabular}[c]{@{}l@{}}C4(\%)\\ (n=39)\end{tabular} &
  \begin{tabular}[c]{@{}l@{}}C1(\%)\\ (n=113)\end{tabular} &
  \begin{tabular}[c]{@{}l@{}}C4(\%)\\ (n=37)\end{tabular} &
  \begin{tabular}[c]{@{}l@{}}C1(\%)\\ (n=162)\end{tabular} &
  \begin{tabular}[c]{@{}l@{}}C4(\%)\\ (n=46)\end{tabular} &
  \begin{tabular}[c]{@{}l@{}}C1(\%)\\ (n=152)\end{tabular} &
  \begin{tabular}[c]{@{}l@{}}C4(\%)\\ (n=71)\end{tabular} &
  \begin{tabular}[c]{@{}l@{}}C1(\%)\\ (n=101)\end{tabular} &
  \begin{tabular}[c]{@{}l@{}}C4(\%)\\ (n=40)\end{tabular} &
  \begin{tabular}[c]{@{}l@{}}C1(\%)\\ (n=162)\end{tabular} \\
Vulnerable &
  28 &
  6 &
  24 &
  5 &
  20 &
  5 &
  29 &
  3 &
  22 &
  4 &
  40 &
  15 &
  28 &
  7 \\
\begin{tabular}[c]{@{}l@{}}English Not\\ Primary Language\end{tabular} &
  31 &
  15 &
  25 &
  16 &
  10 &
  14 &
  44 &
  15 &
  35 &
  12 &
  30 &
  12 &
  34 &
  15 \\
Indigenous &
  40 &
  10 &
  28 &
  10 &
  28 &
  11 &
  69 &
  7 &
  33 &
  8 &
  33 &
  7 &
  19 &
  13 \\
No Preschool &
  24 &
  11 &
  22 &
  12 &
  22 &
  13 &
  22 &
  11 &
  26 &
  12 &
  21 &
  10 &
  20 &
  15 \\
Remoteness – Cities &
  23 &
  76 &
  45 &
  66 &
  51 &
  64 &
  0 &
  77 &
  40 &
  68 &
  35 &
  75 &
  35 &
  70 \\
Remoteness – Regional &
  57 &
  22 &
  48 &
  31 &
  41 &
  33 &
  54 &
  23 &
  46 &
  31 &
  44 &
  24 &
  50 &
  30 \\
Remoteness – Remote &
  20 &
  2 &
  7 &
  3 &
  8 &
  3 &
  8 &
  3 &
  14 &
  1 &
  21 &
  1 &
  15 &
  0 \\
IRSD – Low &
  94 &
  14 &
  71 &
  18 &
  71 &
  26 &
  100 &
  10 &
  86 &
  13 &
  97 &
  13 &
  94 &
  16 \\ \hline
\end{tabular}}
\label{tab:my-table10}
\vspace{6em}
\small\addtolength{\tabcolsep}{-3pt}
\caption{Geographic locations of the most vulnerable cluster (C4) and the least vulnerable cluster (C1).}
\scalebox{1}{
\begin{tabular}{|ll|l|l|l|l|l|l|}
\hline
\multicolumn{2}{|c|}{Physical} &
  \multicolumn{1}{c|}{Social} &
  \multicolumn{1}{c|}{Emotional} &
  \multicolumn{1}{c|}{Language} &
  \multicolumn{1}{c|}{Communication} &
  \multicolumn{1}{c|}{Vuln 1} &
  \multicolumn{1}{c|}{Vuln 2} \\ \hline
\multicolumn{1}{|l|}{C4} &
  \begin{tabular}[c]{@{}l@{}}Far north, very small\\  number central Queensland,\\  small number south east\end{tabular} &
  \begin{tabular}[c]{@{}l@{}}North west, small\\  number central Queensland\end{tabular} &
  \begin{tabular}[c]{@{}l@{}}North west, small\\  number in Central\\  Queensland\end{tabular} &
  Far north &
  \begin{tabular}[c]{@{}l@{}}North west, \\  small number in\\  south west and coastal\\ areas\end{tabular} &
  \begin{tabular}[c]{@{}l@{}}Far north, very\\  small number central\\ coast, small number\\  south east\end{tabular} &
  \begin{tabular}[c]{@{}l@{}}Far north, small \\ number\\ south east\end{tabular} \\
\multicolumn{1}{|l|}{C1} &
  South east and central Queensland &
  \begin{tabular}[c]{@{}l@{}}South east and\\ part of central Queensland\\  few regions in the central west\end{tabular} &
  \begin{tabular}[c]{@{}l@{}}North west, small\\  number in\\   Central Queensland\end{tabular} &
  \begin{tabular}[c]{@{}l@{}}South east  and\\ coastal area\end{tabular} &
  \begin{tabular}[c]{@{}l@{}}South east and\\  central Queensland,\\  few regions in Cairns\end{tabular} &
  \begin{tabular}[c]{@{}l@{}}South east and\\  central  Queensland\end{tabular} &
  \begin{tabular}[c]{@{}l@{}}South east and\\ central Queensland\end{tabular} \\ \hline
\end{tabular}}

\label{tab:my-table100}
\end{sidewaystable*}

\newpage
\section{R shiny}   \label{Shiny}
The following description is based on the explanation given in Moraga book \cite{moraga2019geospatial} chapter thirteen, pages (203-215)
\\
R shiny is a web interactive interface used to build applications in the statistical software package R. Two R scripts are required: a user-interface script called  {\fontfamily{qcr}\selectfont
ui.R} and a server script called {\fontfamily{qcr}\selectfont
server.R}. The user interface script is in control of the application's layout and appearance. The server script contains the R objects as well as the instructions for displaying them. Shiny applications support interactivity by utilising a feature known as reactivity. Users can enter text, select dates, or change other inputs in this manner, and the R objects displayed will change automatically.\\
The steps below can be used to create reactive objects. Reactive expressions let you control which parts of the application (app) update, and prevent unnecessary computation that can slow down the app). R objects are first added to the user interface. This is accomplished by including output functions in the  {\fontfamily{qcr}\selectfont
ui.R} script that convert R objects to output. Following that, the R code for creating the objects is provided in the {\fontfamily{qcr}\selectfont
server.R
} This script includes an unnamed function as well as two list-like objects called output and input. Input stores the current values of the objects in the application, while output contains all of the instructions for building the R objects. The objects are created with a render function and saved in the output list. By including an input value in a {\fontfamily{qcr}\selectfont render*
} expression, reactivity is created.\\
To make a reactive plot, for example, we need to include a {\fontfamily{qcr}\selectfont
plotOutput} function in the {\fontfamily{qcr}\selectfont
ui.R}  The plot is then created using a {\fontfamily{qcr}\selectfont
renderPlot}  function and added to the output object {\fontfamily{qcr}\selectfont
server.R} \cite{moraga2017spatialepiapp}.


\begin{figure*}[!htbp]
  \centering
  \includegraphics[scale=.8]{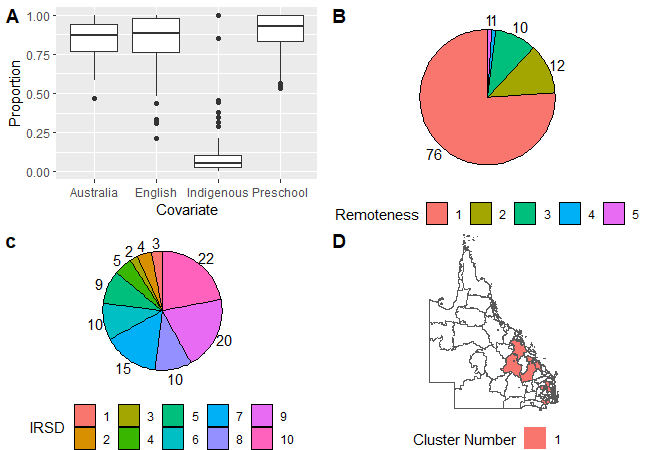}
  \caption{First cluster for Physical using $K$-means, SA2's size=126 (lowest SA2 proportion of Physical vulnerability)}
  \label{fig:sub}

  \centering
  \includegraphics[scale=.8]{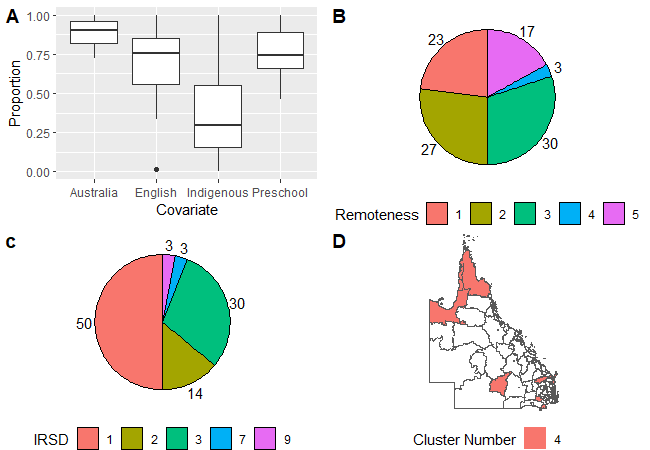}
  \caption{Fourth cluster for Physical using $K$-means, SA2's size=30 (highest SA2 proportion of Physical vulnerability)}
  \label{fig:sub222}
\end{figure*}

\newpage

\begin{figure*}[!htbp]
  \centering
  
  \includegraphics[scale=0.8]{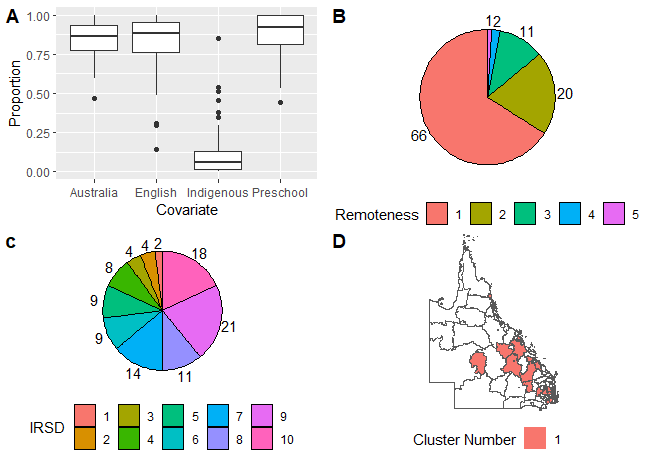}
  \caption{First cluster for Social using $K$-means, SA2's size=111 (lowest SA2 proportion of Social vulnerability)}
  \label{fig:sub11}

  \centering
  
  \includegraphics[scale=0.8]{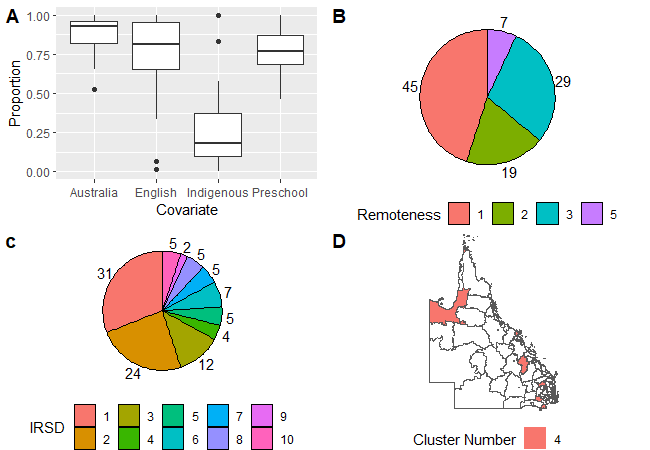}
 \caption{Fourth cluster for Social using $K$-means, SA2's size=42 (highest SA2 proportion of  Social vulnerability)}
  \label{fig:sub22}
\end{figure*}

\newpage
\begin{figure*}[!htbp]

  \centering
  \includegraphics[scale=0.8]{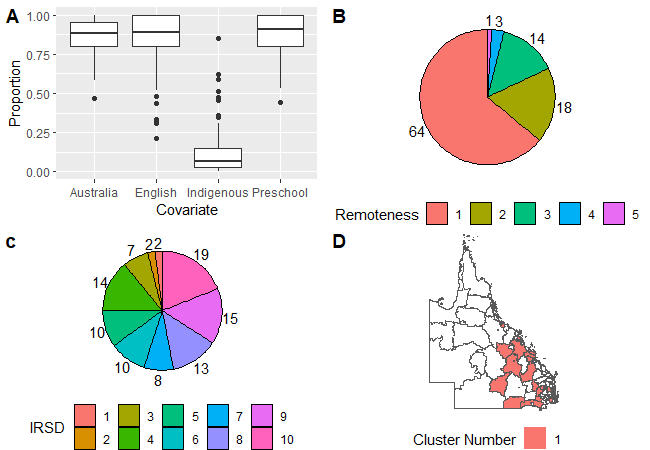}
  \caption{First cluster for Emotional using $K$-means, SA2's size=113 (lowest SA2 proportion of Emotional vulnerability)}
  \label{fig:sub1111}
  \centering
  \includegraphics[scale=0.8]{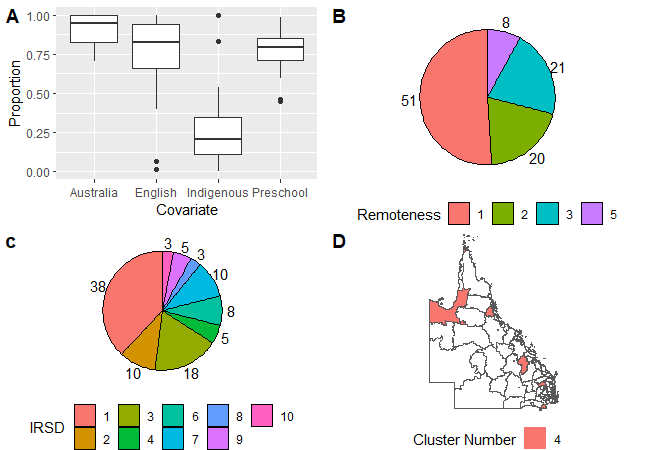}
  \caption{Fourth cluster for Emotional using $K$-means, SA2's size=39 (highest SA2 proportion of Emotional vulnerability)}
  \label{fig:sub10}
\end{figure*}

\newpage

\centering
\begin{figure*}[!htbp]
  \centering
  \includegraphics[scale=0.8]{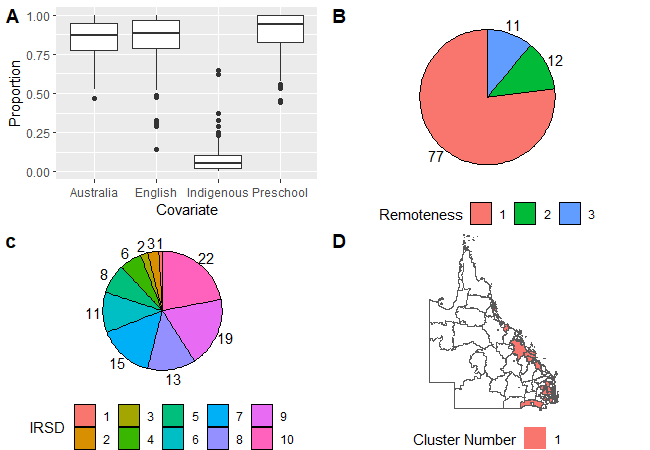}
  \caption{First cluster for Language  using $K$-means, SA2's size=162 (lowest SA2 proportion of Language vulnerability)}
  \label{fig:su}

  \centering
  \includegraphics[scale=0.8]{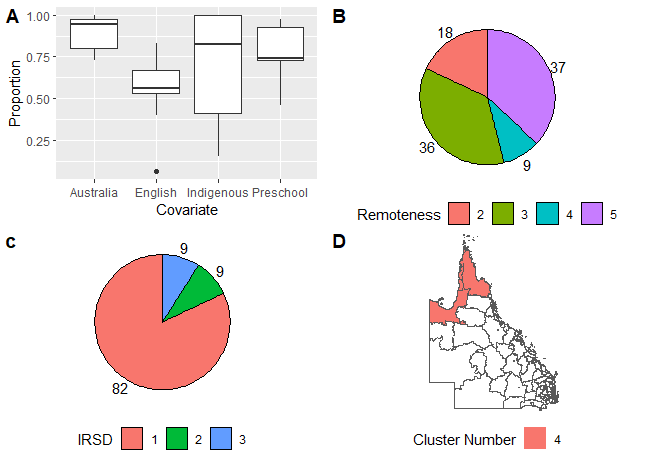}
  \caption{Fourth cluster for Language using K-means, SA2's size=37,  (highest SA2 proportion of Language vulnerability)}
  \label{fig:s}
\end{figure*}

\newpage
\begin{figure*} [!htbp]
  \centering
  \includegraphics[scale=0.8]{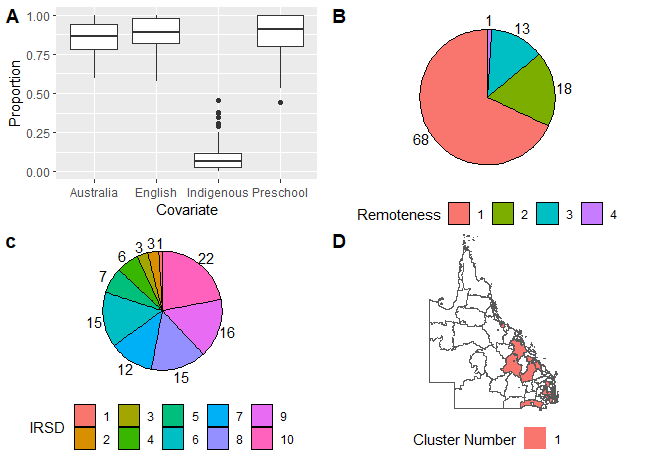}
  \caption{First cluster for Communication using $K$-means, SA2's size=152 (lowest SA2 proportion of Communication vulnerability}
  \label{fig:sub3}
  \centering
  \includegraphics[scale=0.8]{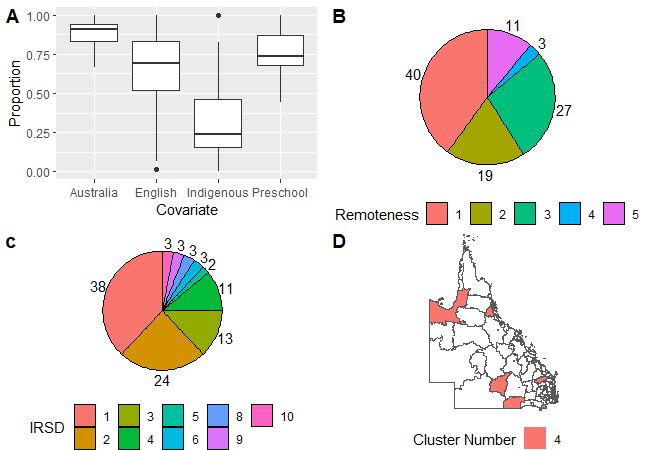}
  \caption{Fourth cluster for Communication using $K$-means, SA2's size=46 (highest SA2 proportion of Communication vulnerability)}
  \label{fig:sub4}
\end{figure*}

\newpage

\begin{figure*}[!htbp]
  \centering
  \includegraphics[scale=0.8]{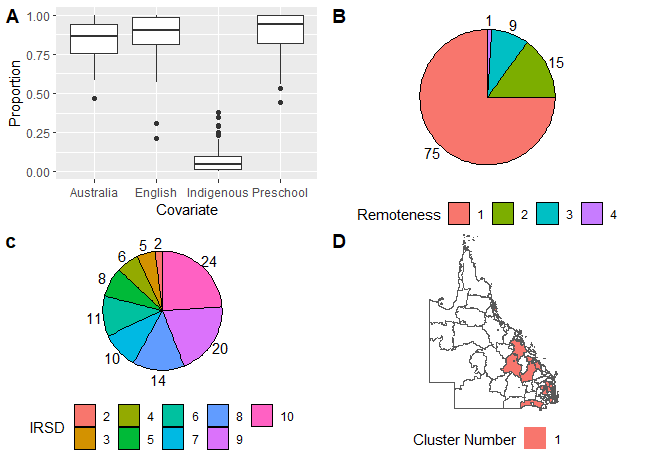}
  \caption{First cluster for Vuln 1 using $K$-means, SA2's size=101 (lowest SA2 proportion of Vulnerability on two or more domain/s)}
  \label{fig:sub6}

  \centering
  \includegraphics[scale=0.8]{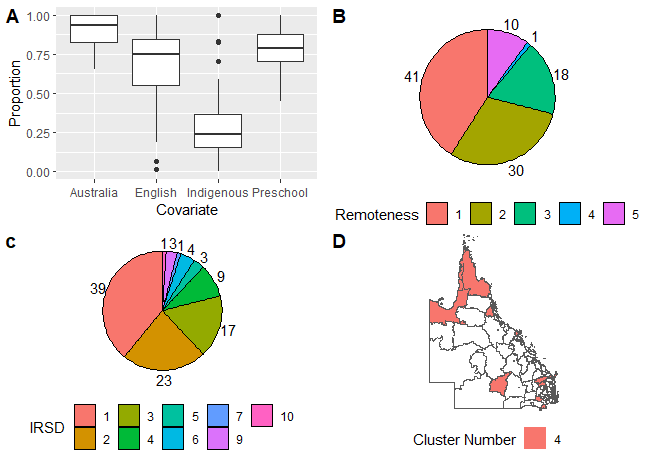}
  \caption{Fourth cluster for Vuln 1 using $K$-means, SA2's size=71  (highest SA2 proportion of Vulnerability on one or more domain/s)}
  \label{fig:sub7}
\end{figure*}

\newpage
\begin{figure*}[!htbp]

  \centering
  \includegraphics[scale=0.8]{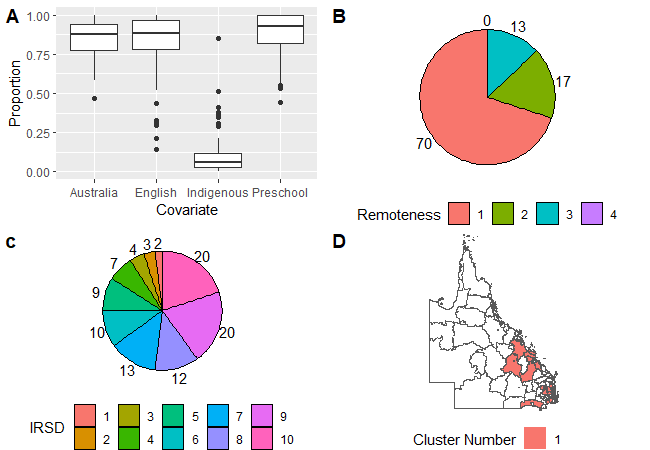}
  \caption{ First cluster for Vuln 2 using $K$-means, SA2's size=162 (lowest SA2 proportion of Vulnerability on two or more domains)}
  \label{fig:sub8}

  \centering
  \includegraphics[scale=0.8]{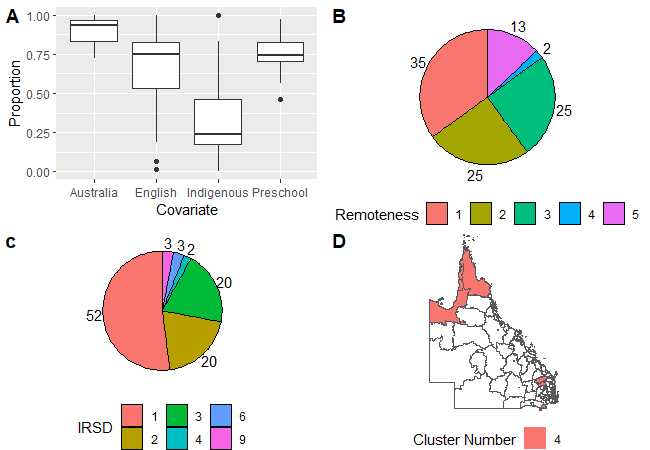}
  \caption{Fourth cluster for Vuln 2 using $K$-means, SA2's size=40 (highest SA2 proportion of Vulnerability on two or more domains)}
  \label{fig:sub9}
\end{figure*}

\newpage
\section{Results from each type of AEDC domain}  \label{App} 
\begin{table*}[!htbp]
\caption{Summary for the most vulnerable cluster (C4) to other clusters in each AEDC domain. }
\scalebox{0.6}{
 \begin{tabular}{llllllll}  \\ \hline
Domain                                                              & Physical                                                                                                       & Social                                                                                     & Emotional                                                                                   & Language  & Communication                                                                                       & Vuln 1                                                                                                        & Vuln 2 \\                      \hline
 Size                                                                & 30                                                                                                             & 42                                                                                         & 39                                                                                          & 37        & 46                                                                                                  & 71                                                                                                            & 40                           \\
 Australia                                                           & (1-3)\%                                                                                                        & (1-4)\%                                                                                    & (1-4)\%                                                                                     & (1-3)\%   & (1-3)\%                                                                                             & (2-5)\%                                                                                                       & (1-5)\%                      \\
English                                                             & (14-16)\%                                                                                                      & (7-8)\%                                                                                    & (1-5)\%                                                                                     & (23-29)\% & (12-23)\%                                                                                           & (11-18)\%                                                                                                     & (15-18)\%                    \\
 Indigenous                                                          & (19-30)\%                                                                                                      & (9-18)\%                                                                                   & (9-17)\%                                                                                    & (42-62)\% & (8-24)\%                                                                                            & (14-26)\%                                                                                                     & (6-12)\%                     \\
 Preschool                                                           & (2-13)\%                                                                                                       & (3-10)\%                                                                                   & (4-9)\%                                                                                     & (2-4)\%   & (4-13)\%                                                                                            & (1-12)\%                                                                                                      & (1-10)\%                     \\
 \begin{tabular}[c]{@{}l@{}}Remoteness\\ (very remote)\end{tabular}  & 17\%                                                                                                           & 7\%                                                                                        & 8\%                                                                                         & 37\%      & 11\%                                                                                                & 18\%                                                                                                          & 13\%                         \\
 \begin{tabular}[c]{@{}l@{}}IRSD (most\\ disadvantaged)\end{tabular} & 94\%                                                                                                           & 71\%                                                                                       & 71\%                                                                                        & 100\%     & 86\%                                                                                                & 97\%                                                                                                          & 94\%                         \\
 \begin{tabular}[c]{@{}l@{}}Geographic\\  distribution\end{tabular}  & \begin{tabular}[c]{@{}l@{}}Far north, very small \\ number central  coast, \\ small number south east\end{tabular} & \begin{tabular}[c]{@{}l@{}}North west, small \\ number in  Central\\ Queensland\end{tabular} & \begin{tabular}[c]{@{}l@{}}North west, small  \\number in  Central \\Queensland\end{tabular} & Far north & \begin{tabular}[c]{@{}l@{}}North west, small\\  number in south \\west and coastal \\areas\end{tabular} & \begin{tabular}[c]{@{}l@{}}Far north, very\\ small number central\\ coast, small number\\ south east\end{tabular} & \begin{tabular}[c]{@{}l@{}}Far north, \\small number \\ south east \end{tabular} \\ \hline
 \end{tabular}}
 \label{tab:my-tablesummary}
\end{table*}
Table \ref{tab:my-tablesummary} provides a summary of the most vulnerable cluster (C4) to other clusters in each type of AEDC domain.
 For the Physical health domain, the cluster sizes for the least (C1) to most (C4) proportions of vulnerable children were 126, 219, 151 and 30 SA2 areas, respectively. For the most physically vulnerable cluster, there were roughly similar percentages of children from inner cities and regional areas (23-30\%), in comparison to the other clusters where most children were from inner cities (76\%); there was a notably higher percentage of children from very remote areas (17\%) compared to the other clusters ($<$ 2\%); 94\% were in the lowest four rungs of socioeconomic disadvantage (most disadvantaged), which is substantially more than the other clusters (14\%-61\%); there was a lower percentage of children (15\% less) with English as their first language; less children went to preschool (2-13\% less); and there was a higher proportion of Indigenous children (19-30\%). These SA2 areas in (C4) were located in the north of Queensland and a small number were also identified in the central coast and south east of Queensland.\\
 For the social health domain, the cluster sizes for the least (C1) to most (C4) proportions of vulnerable children were 168, 205, 111 and 42 SA2 areas, respectively. For the most vulnerable cluster in this domain, there were higher percentages of children from inner cities (45\%), in comparison to the other clusters; there was a higher percentage of children from very remote areas (7\%) compared to the other clusters ($<$ 3\%); 71\% were in the lowest four rungs of socioeconomic disadvantage (most disadvantaged), which is substantially more than the other clusters (18\%-51\%); there was a lower percentage of children (8\% less) with English as their first language; less children went to preschool (3-10\% less); and there was a higher proportion of Indigenous children (9-18\%). These SA2 areas in (C4) were located in the north west of Queensland and a small number were also identified in the central Queensland of Queensland.
 \\				
 For the emotional health domain, the cluster sizes for the least (C1) to most (C4) proportions of vulnerable children were 1113, 194, 180 and 39 SA2 areas, respectively, For  the most vulnerable cluster in this domain, there were high percentages of children from inner cities and regional areas (20-51\%), in comparison to the other clusters where most children were from inner cities; there was a notably higher percentage of children from very remote areas (8\%) compared to the other clusters ($<$ 3\%); 71\% were in thee bottom four rungs of socioeconomic disadvantage (most disadvantaged), which is substantially more than the other clusters (26\%-38\%); there was a lower percentage of children  (~5\% less) with English as their first language; less children went to preschool (4-9\%); and there was a higher proportions of Indigenous children (9-17\%). These SA2 areas in (C4) were located in located in the far north of Queensland.
  \\
  For the language health domain, the cluster sizes for the least (C1) to most (C4) proportions of vulnerable children were 162, 194, 133, and 37 SA2 areas, respectively. For the most vulnerable cluster in this domain, there were low percentages of children from inner cities and regional areas (0-18\%), in comparison to the other clusters where most children were from regional areas; there was a notably higher percentage of children from very remote areas (36\%) compared to the other clusters ($<$ 8\%); 100\% were in the bottom four rungs of socioeconomic disadvantage (most disadvantaged), which is substantially more than the other clusters (10\%-78\%); there was a notably higher percentage of children  (23-29\% less) with English as their first language; less children went to preschool (2-4\%); and there was a notably higher proportion of Indigenous children (42-62\%) . These SA2 areas in (C4) were located  in the far north of Queensland.
 \\
  For the communication health domain, the cluster sizes for the least (C1) to most (C4) proportions of vulnerable children were 152, 195, 133, and 46 SA2 areas, respectively. For the most vulnerable cluster in this domain, there were roughly higher percentages of children from inner cities and regional areas (27-40\%), in comparison to the other clusters where most children were from inner cities; there was a notably higher percentage of children from very remote areas (11\%) compared to the other clusters ($<$ 7\%);86\% were in the bottom four rungs of socioeconomic disadvantage (most disadvantaged), which is substantially more than the other clusters (13\%-67\%); there was a notably higher percentage of children  (12-23\%) with English as their first language; less children went to preschool (4-13\%); and there was a notably higher proportion of Indigenous children (8-42\%). These SA2 areas in (C4) were located in the north west of Queensland and a small number also identified in the south west and coastal areas of Queensland.
 \\
  For the Vuln 1 indicator, the cluster sizes for the least (C1) to most (C4) proportions of vulnerable children were 101, 181, 173, and 71 SA2 areas, respectively. For the most vulnerable cluster in this indicator, there were roughly higher percentages of children from inner cities and regional areas (23-35\%), in comparison to the other clusters where most children were from inner cities; there was a notably higher percentage of children from very remote areas (18\%) compared to the other clusters ($<$ 3\%); 97\% were in the bottom four rungs of socioeconomic disadvantage (most disadvantaged), which is substantially more than the other clusters (13\%-53\%); there was a notably higher percentage of children (11-18\%) with English as their first language; less children went to preschool (1-12\%); and there was a notably higher proportion of Indigenous children (14-26\%). These SA2 areas in (C4) were located in the far north of Queensland and a small number also identified in the south east and central coast of Queensland.
\\
  With regard to Vuln 2 indicator, the cluster sizes for the least (C1) to most (C4) proportions of vulnerable children  were 162, 207, 117, and 40, SA2 areas, respectively. For the most Vuln 2 cluster: there were roughly higher percentages of children from inner cities and regional areas (35-25\%), in comparison to the other clusters where most children were from inner cities; there was a notably higher percentage of children from very remote areas (13\%) compared to the other clusters ($<$ 4\%); 94\% were in thee bottom four rungs of socioeconomic disadvantage (most disadvantaged), which is substantially more than the other clusters (16\%-46\%); there was a notably higher percentage of children  (15-18\%) with English as their first language; less children went to preschool (1-10\%); and there was a higher proportion of Indigenous children (6-12\%), The most Vuln 2 cluster domain can be found in far north of Queensland and a small number also identified in south east. Compared to Vuln 1, the geographic distribution is less in central coast and south east. there was a slightly higher percentage of children (3\%) with English as their first language; less children went to preschool (2\% ); and there was a higher proportion of Indigenous children (8-14\%).\\

\end{appendix}
\end{document}